\documentclass[12pt]{article} 

\usepackage{amsfonts,amsmath,epsfig,epsfig,cite}

\newcommand{\ii}{\textrm{i}}
\newcommand{\ee}{\textrm{e}}
\renewcommand{\vec}[1]{\boldsymbol{#1}}
\newcommand{\be}{\begin{equation}}
\newcommand{\en}{\end{equation}}

\begin{document}
\title{Love waves in a \\piezoelectric layered structure}

\author{Bernard Collet and Michel Destrade} 
\date{2006}

\maketitle 

\begin{abstract}

Consider a  layer made of a m3m insulator crystal, 
with faces cut parallel to a symmetry plane.
Then bond it onto a semi-infinite mm2 piezoelectric substrate. 
For a $X$- or $Y$-cut of the substrate, a Love wave can propagate in 
the resulting structure and 
the corresponding dispersion equation is derived analytically. 
It turns out that a fully explicit treatment can also be conducted in 
the case of a $Y$-cut rotated about $Z$. 
In the case of a germanium layer over a potassium niobate substrate,
the wave exists at any wavelength for  $X$- and $Y$-cuts but this 
ceases to be the case for rotated cuts, with the appearance of 
forbidden ranges. 
By playing on the cut angle, the Love wave can be made to travel 
faster than, or slower than, or at the same speed as, the shear bulk 
wave of the layer.  
A by-product of the analysis is the derivation of the explicit secular 
equation for the Bleustein-Gulyaev wave in the substrate alone, 
which corresponds to an asymptotic behavior of the Love wave. 

\end{abstract}
%

\section{Introduction} 

Layered structures, 
especially film/coating substrate systems,
play an important role in micro-electro-mechanical systems (MEMS) 
and in microelectronics packages. 
In order to achieve high performance, many surface acoustic wave (SAW) 
devices/sensors are made of layered architectures such as, 
for instance, a dielectric, or a piezoelectric, or a non-piezoelectric 
semiconductor (finite-thickness) layer deposited onto a 
(semi-infinite) piezoelectric substrate. 
For certain configurations, it is possible to have a one-component 
wave travel in the structure, in the direction of the interface: this 
guided (shear-horizontal) Love wave leaves the upper face of the layer
 free of mechanical 
tractions, its amplitude varies sinusoidally through the thickness of 
the layer and then decays rapidly with depth in the substrate, and it 
is such that all fields are continuous at the layer/substrate 
interface.
      
Love waves in piezoelectric layered acoustic devices are most suitable 
for high frequency filters because of their high phase velocity,
and they also show great promise in bio-sensors applications with 
liquid environments because of their high sensitivity. 
Consequently, they have received much attention over the years. 
For example, Lardat et al. \cite{LaMT71} found under which conditions 
a piezoelectrically stiffened Love wave exist and presented 
experimental and analytical results on surface wave delay lines. 
Kessenikh et al. \cite{KeLS82} investigated surface Love waves 
in piezoelectric substrates of classes 6, 4, 6mm, 4mm, 622, and 422 
with an isotropic dielectric layer.
Hanhua and Xingjiao \cite{HaXi93} studied Love waves for 
a structure made of a 6mm piezoelectric layer and a 6mm piezoelectric 
substrate, with a common symmetry axis in the plane of the interface, 
while Darinskii and Weihnacht \cite{DaWe01} had a similar structure, 
made of 2mm piezoelectric layer and substrate with common symmetry 
axes, one of which is aligned with the propagation direction and 
another is aligned with the normal to the interface.
Jakoby and Vellekoop \cite{JaVe97} reviewed the properties of Love 
waves and associated numerical methods for a 
piezoelectric/piezoelectric layered structure, when the substrate is 
made of ST-cut quartz; 
so did Ogilivy \cite{Ogil97}, with special emphasis on the 
mass-sensitivity loading of biosensors.
The reader can find additional pointers to the literature on Love 
waves for piezoelectric sensors in those articles and in the references
therein, as well as in the reviews by Farnell and Adler \cite{FaAd72}
or by Gulyaev \cite{Guly98}.

\begin{figure}
 \centering 
  \epsfig{figure=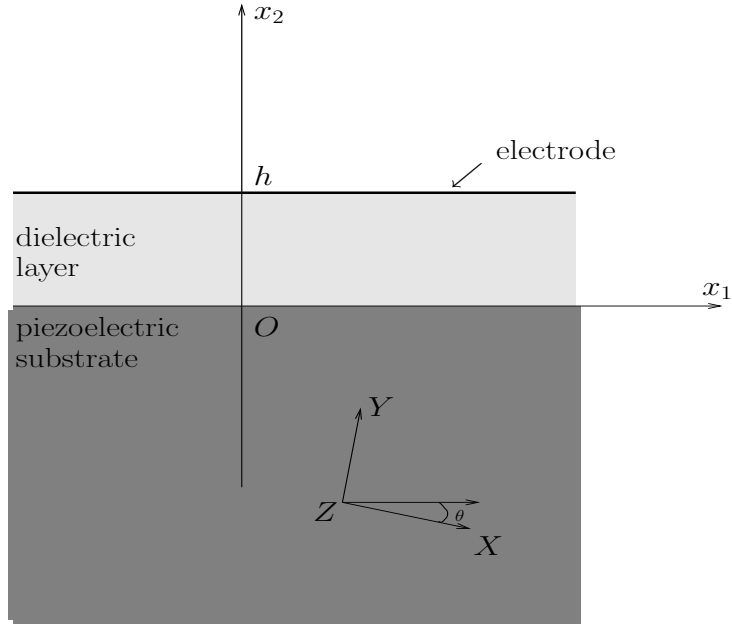, height=.6\textwidth, width=.7\textwidth}
\caption{{\small \sffamily{Geometry of the layered structure}}}
\label{Figure1}
\end{figure}

The present work is concerned with the propagation of Love waves in a 
composite structure, made of a m3m cubic insulator
(non-piezoelectric semiconductor) layer of finite thickness $h$, 
bonded onto a $Y$-cut rotated about $Z$, mm2 orthorhombic, 
piezoelectric substrate, see Fig.\ref{Figure1}. 
With this singular structure, it is possible to exploit the 
potentiality offered by an efficient direct analytic method. 
The steps to be followed are: 
\textit{(i)} Determine the general solution for the layer, satisfying 
the boundary conditions on the upper face; 
\textit{(ii)} Derive some fundamental equations in the substrate 
using the Stroh formulation of the equations of motion there; 
\textit{(iii)} Match the solutions at the layer/substrate interface; 
and 
\textit{(iv)} Deduce the dispersion equation for piezoelectric 
Love waves in explicit form. 

The dispersion relation admits a denumerable number of solutions
(bran\-ches) and the solutions for the mechanical displacements, shear 
stresses, electric potentials, and electric inductions are deduced  
explicitly in the case of a metalized, mechanically free, upper 
surface, brought to the zero electric potential (short-circuit).
In passing, the explicit secular equations for the Bleustein-Gulyaev 
wave speed and for the limiting wave speed 
(substrate only, no overlayer) are found as a cubic and as a sextic 
in the squared wave speed respectively, for rotated cuts.
The special cases of a $X$-cut or $Y$-cut are treated separately. 
The effects of the angle of the cut on the phase velocity of the first 
modes are illustrated numerically for a
specific layered structure, namely a germanium layer over a 
potassium niobate substrate and the appearance of a forbidden band of 
frequency is uncovered for a rotated cut, in sharp contrast with the 
non-rotated cuts where the waves exist for all frequencies.

\section{The layer} 

First consider the upper layer, which is made of a cubic, m3m 
non-piezoelectric (semi-conductor) crystal, with mass density 
$\hat{\rho}$.
For two-dimensional motions (independent of $x_3$), 
the anti-plane equation of motion decouples from its in-plane 
counterpart and reads ($0 \le x_2 \le h$)
\begin{equation} \label{motion3}
  \hat{c}_{44} (\hat{u}_{3,11} + \hat{u}_{3,22}) 
    = \hat{\rho} \hat{u}_{3,tt},
\end{equation}
where $\hat{u}_3$ is the anti-plane mechanical 
displacement, $\hat{c}_{44}$ the transverse 
stiffness, and the comma denotes partial differentiation. 
For a solution in the form of an inhomogeneous wave traveling with 
speed $v$ and wave number $k$ in the $x_1$ direction, such as 
\begin{equation} \label{u3}
  \hat{u}_3 = \hat{U}_3(kx_2) \ee^{\ii k(x_1 - vt)},
\end{equation}
(where $\hat{U}_3$ is a function of $kx_2$ alone),
the equation of motion \eqref{motion3} reduces to
\begin{equation}
  \hat{U}_3'' + \left(\dfrac{v^2}{\hat{v}^2} - 1 \right)
   \hat{U}_3 =0, \quad \text{ where } \quad 
    \hat{v} := \sqrt{\hat{c}_{44} / \hat{\rho}}.
\end{equation}
The general solution to this second-order differential equation 
is either \textit{(i)}
\begin{multline} \label{U_3gnl(i)}
  \hat{U}_3 (kx_2) = 
    \hat{U}_3(0)
      \left(
         \cos\sqrt{\frac{v^2}{\hat{v}^2} - 1} k x_2  \right. \\
          \left. + A  \sin\sqrt{\frac{v^2}{\hat{v}^2} - 1} k x_2
            \right),
            \quad \text{when} \quad 
    v>\hat{v}, 
\end{multline}
or \textit{(ii)}
\begin{multline} \label{U_3gnl(ii)}
  \hat{U}_3 (kx_2) = 
    \hat{U}_3(0)
      \left(
         \cosh\sqrt{1-\frac{v^2}{\hat{v}^2}} k x_2 \right. \\
           \left. + A  \sinh\sqrt{1-\frac{v^2}{\hat{v}^2}} k x_2
            \right),
            \quad \text{when} \quad 
    v<\hat{v}, 
\end{multline}
where $A$ is a constant to be determined from the boundary condition 
at $x_2 = h$. 
This latter condition is that the upper face of the layer be free 
of mechanical tractions, so that 
$\hat{\sigma}_{23} = \hat{c}_{44} \hat{u}_{3,2}$ 
is zero there. 
Then it follows from \eqref{U_3gnl(i)} and 
\eqref{U_3gnl(ii)} that: 
$A = \tan \sqrt{v^2/\hat{v}^2 - 1} kh$ in 
Case \textit{(i)}, and that: 
$A = -\tanh \sqrt{1 - v^2/\hat{v}^2} kh$ in 
Case \textit{(ii)}. 
Consequently the mechanical field is now entirely known in the layer.

The procedure to find the electrical field is similar, 
and even simpler because the layer is not piezoelectric. 
Taking the electric potential $\hat{\phi}$ in the form
\begin{equation} \label{phi}
  \hat{\phi} = \hat{\varphi}(kx_2) \ee^{\ii k(x_1 - vt)},
\end{equation}
(where $\hat{\varphi}$ is a function of $kx_2$ alone), 
Poisson's equation: $\Delta \hat{\phi} = 0$ reduces to 
\begin{equation}
  \hat{\varphi}'' -  \hat{\varphi} =0, 
\end{equation}
with general solution 
\begin{equation} \label{phiGnl}
  \hat{\varphi} (kx_2) = 
      \hat{\varphi}(0)
       \left(  \cosh k x_2 + B  \sinh k x_2 \right),
\end{equation}
where $B$ is a constant to be determined from the boundary condition 
on the upper face of the layer $x_2 = h$. 
A thin metallic film covers that face, and it is grounded to potential 
zero: $\hat{\varphi}(kh)=0$. 
It then follows from \eqref{phiGnl} that 
$B = - \coth kh$, 
and consequently, that the electrical field is now 
entirely known in the layer.

For the problem at hand, only the values of the fields at the interface
$x_2=0$ between the layer and the substrate are needed. 
Because the mechanical displacement $\hat{u}_3$ and the electrical 
potential $\hat{\phi}$ are in the forms \eqref{u3} and \eqref{phi}, 
respectively, the mechanical traction $\hat{\sigma}_{23}$ and the 
electrical displacement $\hat{D}_2$ must also be of a similar form, 
due to the constitutive relations: 
$\hat{\sigma}_{23} = \hat{c}_{44} \hat{u}_{3,2}$ 
and $\hat{D}_2 = -\hat{\epsilon}_{11} \phi_{,2}$, where 
$\hat{\epsilon}_{11}$ is the dielectric constant of the layer. 
Hence, 
\begin{equation}
   \hat{\sigma}_{23} = \ii k \hat{t}_{23}(kx_2) \ee^{\ii k(x_1 - vt)},
	 \quad 
	 \hat{D}_2 = \ii k \hat{d}_2(kx_2) \ee^{\ii k(x_1 - vt)},
\end{equation}
(say), where $\hat{t}_{23}$ and $\hat{d}_2$ are functions of $kx_2$ 
alone.
In particular, the conclusions drawn from 
the calculations conducted above 
are that at the layer/sub\-stra\-te interface, 
\be \label{layerBelow}
\hat{t}_{23}(0) = - \ii \hat{c} \hat{U}_3(0), 
\quad 
\hat{d}_2(0)  = - \ii \hat{\epsilon} \hat{\varphi}(0),
\en
where 
\begin{align} \label{epsilonHat}
& \hat{c} := 
  \hat{c}_{44}
      \sqrt{\dfrac{v^2}{\hat{v}^2} - 1} 
         \tan \sqrt{\dfrac{v^2}{\hat{v}^2} - 1} kh, 
\quad \text{when} \quad  v > \hat{v},
  \notag \\
& \hat{c} := 
   -\hat{c}_{44}
     \sqrt{1 - \dfrac{v^2}{\hat{v}^2}}
      \tanh \sqrt{1 - \dfrac{v^2}{\hat{v}^2}} kh, 
\quad \text{when} \quad v < \hat{v},
  \notag \\
& \hat{\epsilon}:= \hat{\epsilon}_{11} \coth kh. 
\end{align}

\section{The substrate} 

The substrate occupies the half-space $x_2 \le 0$ and 
is made of the superstrong piezoelectric crystal, 
potassium niobate KbNO$_3$, with orthorhombic mm2 symmetry.
Cut the crystal along a plane containing the $Z$ axis and making an 
angle $\theta$ with the $XY$ plane.
Let $x_1x_2x_3$ be the coordinate system obtained after the rotation 
\begin{equation}
  \begin{bmatrix}
    m & n & 0 \\ -n & m & 0 \\ 0 & 0 & 1 
  \end{bmatrix}, \quad \text{where}
\quad m = \cos \theta, \quad n= \sin \theta.
\end{equation}
Yet again, for a two-dimensional motion 
(independent of $x_3$) the anti-plane strain and stress 
decouple from their in-plane counterparts. 
Hence, with $u_1 = u_2 =0$, $u_3 = u_3(x_1, x_2, t)$ for the 
mechanical displacement, and $\phi = \phi(x_1, x_2, t)$ for the 
electric potential, the constitutive relations yield 
$\sigma_{11} = \sigma_{22} = \sigma_{33} = \sigma_{12} = 0$ for 
the stress components and $D_3 =0$ for the electrical displacement, 
and they reduce to 
\begin{align}\label{constitutive}
  & \sigma_{23} = 
  c_{44} u_{3,2} + c_{45}u_{3,1} + e_{14}\phi_{,1} + e_{24}\phi_{,2},
\notag \\
  & \sigma_{31} = 
  c_{45} u_{3,2} + c_{55}u_{3,1} + e_{15}\phi_{,1} + e_{14}\phi_{,2},
\notag \\
  & D_1 = 
  e_{14} u_{3,2} + e_{15}u_{3,1} 
     - \epsilon_{11}\phi_{,1} - \epsilon_{12}\phi_{,2},
\notag \\
  & D_2 = 
  e_{24} u_{3,2} + e_{14}u_{3,1} 
     - \epsilon_{12}\phi_{,1} - \epsilon_{22}\phi_{,2},
\end{align}
where the $c_{ij}$, $e_{ij}$, $\epsilon_{ij}$ are related to the 
corresponding quantities $\tilde{c}_{ij}$, $\tilde{e}_{ij}$, 
$\tilde{\epsilon}_{ij}$ in the crystallographic coordinate system 
$XYZ$ through tensor transformations \cite{Auld73} as:
\begin{align} \label{c_e_epsilon}
  &  c_{44} = m^2 \tilde{c}_{44} + n^2 \tilde{c}_{55}, 
  && c_{55} = n^2 \tilde{c}_{44} + m^2 \tilde{c}_{55}, 
 \notag \\
  &  c_{45} = mn (\tilde{c}_{44} - \tilde{c}_{55}),
  && e_{24} = m^2 \tilde{e}_{24} + n^2 \tilde{e}_{15}, 
   \notag \\
  &  e_{15} = n^2 \tilde{e}_{24} + m^2 \tilde{e}_{15}, 
  && e_{14} = mn (\tilde{e}_{24} - \tilde{e}_{15}),
   \notag \\
  &  \epsilon_{11} = 
        m^2 \tilde{\epsilon}_{11} + n^2 \tilde{\epsilon}_{22}, 
  && \epsilon_{22} = 
        n^2 \tilde{\epsilon}_{11} + m^2 \tilde{\epsilon}_{22},
  \notag \\
  &  \epsilon_{12} = 
        mn (\tilde{\epsilon}_{22} - \tilde{\epsilon}_{11}).
\end{align}

Now consider the $x_2$-cut, $x_1$-propagation of a Shear-Horizontal
interface acoustic wave that is, a motion with speed $v$ and wave 
number $k$ where the displacement field $u_3$ and the electric 
potential $\phi$ are of the form, 
\begin{equation} \label{uPhi}
 \{ u_3, \phi \} (x_1, x_2, t) 
 = \{ U_3(k x_2), \varphi(k x_2) \} \ee^{\ii k(x_1 - vt)}, 
\end{equation} 
(say), with 
\begin{equation} \label{BC1}
 U_3(-\infty) = 0, \quad \varphi(-\infty) = 0. 
\end{equation}
It follows from the constitutive equations \eqref{constitutive} 
that the traction $\sigma_{32}$ and the electric induction $D_2$ 
are of a similar form, 
\begin{equation} \label{wave2} 
\{ \sigma_{32}, D_2 \} (x_1, x_2, t) =
 \ii k \{ t_{32}(kx_2), d_2(kx_2) \} \ee^{\ii k (x_1 - vt)},
\end{equation}
(say), where 
\begin{equation} \label{BC2}
 t_{32}(-\infty) = 0, \quad d_2(-\infty) = 0. 
\end{equation}

The non-trivial part of the equations of piezoacoustics,  
$\sigma_{ij,j} =  \rho u_{i,tt}$, $D_{i,i} = 0$ 
(where $\rho$ is the mass density of the crystal), 
can be written as a second-order differential system
\cite{MoWe02}:
\begin{equation} \label{2ndOrder}
  T
  \begin{bmatrix}
    U_3'' \\ \varphi''
  \end{bmatrix}
  + 2\ii R 
  \begin{bmatrix}
    U_3' \\ \varphi'
  \end{bmatrix}
  -[Q - (\rho v^2) J]
  \begin{bmatrix}
    U_3 \\ \varphi
  \end{bmatrix}
 = \begin{bmatrix}
    0 \\ 0
  \end{bmatrix},
\end{equation}
where the prime denotes differentiation with respect to $kx_2$ and
\begin{align} \label{TRQJ}
&  T :=  \begin{bmatrix}
           c_{44} &  e_{24} \\ 
           e_{24} &  -\epsilon_{22}  
         \end{bmatrix},
&& R :=  \begin{bmatrix}
           c_{45} &  e_{14} \\ 
           e_{14} & -\epsilon_{12} 
         \end{bmatrix},
\notag \\
&  Q :=  \begin{bmatrix}
            c_{55} &  e_{15} \\   
            e_{15} & -\epsilon_{11} 
         \end{bmatrix},
&& J :=  \begin{bmatrix}
             1 & 0 \\   
             0 & 0 
         \end{bmatrix},
\end{align}
or as a first-order differential system in the form 
\begin{equation}  \label{motion}
\vec{\xi}' = \ii N \vec{\xi}, 
 \quad
 \text{ where }
 \quad
\vec{\xi}(kx_2) := [U_3, \varphi, t_{31}, d_2]^t, 
\end{equation}
and $N$ has the Stroh block structure 
\begin{equation} \label{N}
  N = 
  \begin{bmatrix}
    N_1 & N_2 \\
    N_3 + (\rho v^2) J & N_1^t
  \end{bmatrix},
\end{equation}
with \cite{Ting96}
\begin{equation} \label{Ni}
   N_1 = -T^{-1} R, \quad
   N_2 = T^{-1}, \quad 
   N_3 = R T^{-1} R - Q.
\end{equation}
Here the components of $N$ are
easily computed from \eqref{TRQJ} and \eqref{Ni}, 
by hand or using a Computer Algebra System. 

Seeking a solution to \eqref{2ndOrder} in the form: 
$[U_3, \varphi]^t = [U_3(0), \varphi(0)]^t \ee^{\ii kqx_2}$,
where $q$ is constant, 
yields the \textit{propagation condition}
\begin{equation} \label{quartic}
  \text{det } [q^2 T + 2 q R + Q - (\rho v^2) J] = 0, 
\end{equation}
a quartic in $q$. 
Explicitly,
\begin{multline}
c_{44}^D q^4
 + 2\dfrac{\epsilon_{12}}{\epsilon_{22}}c_{16}^D q^3 
  + (c_{45}^D - \rho v^2)q^2 \\
   + 2\dfrac{\epsilon_{12}}{\epsilon_{22}}(c_{26}^D - \rho v^2)q 
     + \dfrac{\epsilon_{11}}{\epsilon_{22}}(c_{55}^D - \rho v^2)
       = 0,
\end{multline}
where
\begin{align}
 & c_{44}^D = c_{44} + \dfrac{e_{24}^2}{\epsilon_{22}}, 
\quad 
 c_{16}^D = c_{44} + c_{45}\dfrac{\epsilon_{22}}{\epsilon_{12}}
              + 2 \dfrac{e_{15}e_{24}}{\epsilon_{12}}, 
\notag \\
& 
c_{45}^D = c_{55} + c_{44}\dfrac{\epsilon_{11}}{\epsilon_{22}} 
  + 2 \dfrac{e_{15}e_{24}}{\epsilon_{22}} 
   + 4 c_{45}\dfrac{\epsilon_{12}}{\epsilon_{22}} 
    + 4 \dfrac{e_{14}^2}{\epsilon_{22}},
\notag \\
& c_{55}^D = c_{55} + \dfrac{e_{15}^2}{\epsilon_{11}}, 
\quad 
c_{26}^D = c_{55} + c_{44}\dfrac{\epsilon_{11}}{\epsilon_{12}}
              + 2 \dfrac{e_{15}e_{24}}{\epsilon_{12}}. 
\end{align}
Out of the four possible roots, only two have a negative imaginary 
part, insuring exponential decay as $x_2 \rightarrow -\infty$.
Computing these qualifying roots analytically is not an easy matter, 
because here the speed $v$ is still an unknown. 
Although it is actually possible to do so \cite{Fu05, DeFu05}, 
this paper follows a different route and makes extensive use of the 
``fundamental equations'' derived in \cite{Dest03d, Dest04a, Dest05}.  
They read
\begin{equation} \label{fundamental}
\vec{\xi}(0) \cdot M^{(n)} \overline{\vec{\xi}}(0) = 0, 
\quad \text{where} \quad 
 M^{(n)} :=   \begin{bmatrix} 0   & I \\
                             I & 0 
   \end{bmatrix} N^n, 
\end{equation} 
and $n$ is any positive or negative integer. 

\section{The layer/substrate structure} 

The continuity of all fields at the layer/substrate interface $x_2=0$ 
imposes the boundary conditions:
\begin{align} \label{BC}
&  U_3(0) = \hat{U}_3(0),  
&& \varphi(0) = \hat{\varphi}(0), \notag \\
&  t_{23}(0) = \hat{t}_{23}(0),  
&& d_2(0) = \hat{d}_2(0). 
\end{align}
Now the dispersion equations are derived explicitly. 
In the special cases of $X$ and $Y$-cuts, the dispersion equation is 
\textit{exact}: if it is satisfied, then the Love wave exist. 
In the other cases, the  dispersion equation is 
\textit{rationalized}: it has spurious roots, not corresponding to a 
true Love wave so that a subsequent validity check is necessary.

\subsection{Special cases $\theta =0, 90^\circ$}

For a $X$-cut or a $Y$-cut of the substrate, the 
analysis simplifies considerably and a direct treatment 
is possible, leading to an exact dispersion equation, and 
not requiring the use of the fundamental equations. 

When $m=1$, $n=0$, or $m=0$, $n=1$, the parameters 
$c_{45}$, $e_{14}$, $\epsilon_{12}$ vanish according to 
\eqref{c_e_epsilon}. 
Then in \eqref{TRQJ}, $R \equiv 0$ also, and the quartic 
\eqref{quartic} becomes the following biquadratic
\cite{DaWe01}:
\begin{equation} \label{biquadraticZero}
q^4 -Sq^2 + P=0, 
\end{equation}
where the non-dimensional quantities $S$ and $P$ are  
given by 
\begin{align} \label{SP}
-& S = \dfrac{c_{44}\epsilon_{11} + c_{55}\epsilon_{22}
       + 2e_{15}e_{24} -\epsilon_{22}(\rho v^2)}
         {c_{44}\epsilon_{22} + e_{24}^2}, 
\notag \\
& P = \dfrac{c_{55}\epsilon_{11}  
       + e_{15}^2 -\epsilon_{11}(\rho v^2)}
         {c_{44}\epsilon_{22} + e_{24}^2}.
\end{align}

The relevant roots $q_1$ and $q_2$ come in one 
of the two following forms, either 
\begin{multline} \label{qForm}
(a) \; q_1 = -\ii \beta_1, \; q_2 = -\ii \beta_2, 
 \\
\text{or} \quad (b) \; q_1 = -\alpha - \ii \beta, \; 
q_2 = \alpha - \ii \beta,
\end{multline}
where $\beta_i >0$, $\beta>0$.
In either case, $q_1+q_2$ has no real part and a negative 
imaginary part, and $q_1q_2$ is a negative real number. 
Explicitly,
\begin{equation} \label{sp} 
q_1+q_2 = -\ii \sqrt{2\sqrt{P} - S}, \quad 
q_1q_2 = -\sqrt{P}. 
\end{equation}        
The associated eigenvectors $\vec{\zeta^1}$ and $\vec{\zeta^2}$ 
follow from (for instance) the third column of the adjoints to the 
matrices $N - q_1 I$ and $N-q_2I$, as 
$\vec{\zeta^i} = [\vec{a^i}, \vec{b^i}]^t$, ($i=1, 2$)
where 
\be \label{zeta}
\vec{a^i} = \left[q_i^2 + \dfrac{\epsilon_{11}}{\epsilon_{22}},  
                \dfrac{e_{24}}{\epsilon_{22}} q_i^2 
                   + \dfrac{e_{15}}{\epsilon_{22}}\right]^t, \quad 
\vec{b^i} = q_i T\vec{a^i},
\en 
(here $T$ is given by  \eqref{TRQJ}, with components evaluated at 
0$^\circ$ or $90^\circ$).
Then the general solution to the equations of motion is of the form 
\begin{equation} \label{generalSolution}
\vec{\xi}(kx_2) 
 = \gamma_1 \ee^{\ii kq_1 x_2} \vec{\zeta^1} 
 +  \gamma_2  \ee^{\ii kq_2 x_2} \vec{\zeta^2}, 
\end{equation}
where $\gamma_1$, $\gamma_2$ are disposable constants.

At $x_2 = 0$, it can be split as
\be 
[U_3(0), \varphi(0)]^t = A \vec{\gamma}, \quad
[t_{23}(0), d_2(0)]^t = B \vec{\gamma}, 
\en
where
\be
A := [\vec{a^1} |\vec{a^2}], \quad 
B := [\vec{b^1} |\vec{b^2}], \quad 
\vec{\gamma} := [\gamma_1, \gamma_2]^t. 
\en 
Now  the boundary conditions \eqref{BC} and \eqref{layerBelow} give 
the link
\begin{align} \label{exactBC}
 B \vec{\gamma}& = [\hat{t}_{23}(0), \hat{d}_2(0)]^t 
  =  \text{ Diag}(- \ii\hat{c}, - \ii\hat{\epsilon})
                       [\hat{U}_3(0), \hat{\varphi}(0)]^t
 \notag \\
& = -\ii \text{ Diag}(\hat{c}, \hat{\epsilon})A \vec{\gamma},
\end{align}
from which the dispersion equation follows as
\be \label{dispersion}
 | \ii BA^{-1} - \text{Diag}(\hat{c}, \hat{\epsilon})| =0.
\en
It is written in this form to take advantage of the many desirable 
properties of the surface impedance tensor $\ii BA^{-1}$
\cite{AbBa90}; 
thus, this matrix is Hermitian in the subsonic range
(defined below), and of the compact form:
\be
 \ii BA^{-1} = \dfrac{\ii}{q_1+q_2}
    \begin{bmatrix}
     \rho v^2 - c_{55} - c_{44}q_1q_2 & -e_{15} + e_{24}q_1q_2
  \\
     -e_{15} + e_{24}q_1q_2  & \epsilon_{11} - \epsilon_{22}q_1q_2
    \end{bmatrix}.
\en
Moreover, the eigenvalues of the aggregate impedance tensor in 
\eqref{dispersion} are real monotonic decreasing functions of $v$ for 
any fixed $k h$, so that the wave speed of each mode is obtained 
unambiguously from the roots of \eqref{dispersion},
see Shuvalov and Every \cite{ShEv02}. 

Using the identities $q_1^2 + q_2^2 = S$, $q_1^2  q_2^2 = P$ and the 
connections \eqref{SP} and \eqref{sp}, the \textit{exact, explicit, 
dispersion equation} is finally derived as
\be \label{theta=0-90}
 \begin{vmatrix}
 \dfrac{\rho v^2 - c_{55} - c_{44}\sqrt{P}}{\sqrt{2\sqrt{P}-S}} 
    + \hat{c} &  
  - \dfrac{e_{15} + e_{24}\sqrt{P}}{\sqrt{2\sqrt{P}-S}} 
  \\[15pt] 
 - \dfrac{e_{15} + e_{24}\sqrt{P}}{\sqrt{2\sqrt{P}-S}} & 
    \dfrac{\epsilon_{11} + \epsilon_{22}\sqrt{P}}{\sqrt{2\sqrt{P}-S}}
      + \hat{\epsilon}
 \end{vmatrix}
 = 0,
\en
which is a fully explicit  equation, because 
$\hat{c}$, $\hat{\epsilon}$ are defined in \eqref{layerBelow} 
and $S$, $P$ in \eqref{SP}. 

The dispersion equation is valid in the \textit{subsonic range} that 
is, as long as the speed is less than 
the \textit{limiting speed} $v_L$, the smallest speed at which 
the biquadratic \eqref{biquadraticZero} ceases to have 
two roots $q_1$, $q_2$ with a positive imaginary part. 
When $q_1$, $q_2$ are of the form \textit{(a)} in \eqref{qForm}, 
then $v_L$ is root to: $P=0$; 
when they are of the form \textit{(b)}, then $v_L$ is root to: 
$2\sqrt{P} = S$. 
In either case, the wave becomes homogeneous at $v = v_L$ 
because then the roots of the biquadratic are \textit{real}; 
they are: $\pm q$, where $q =  \sqrt{S_L}$ or $q =\sqrt{S_L/2}$,
according to each case  
(here $S_L$ is $S$ given by \eqref{SP} at $v=v_L$).  
The associated eigenvectors are $[\vec{a}, \pm \vec{b}]^t$ where 
\be
\vec{a} = 
 \left[q^2 + \dfrac{\epsilon_{11}}{\epsilon_{22}},  
                \dfrac{e_{24}}{\epsilon_{22}}q^2 
                   + \dfrac{e_{15}}{\epsilon_{22}}\right]^t, \quad 
\vec{b} =  q T\vec{a}.
\en 
The vanishing of the wave away from the interface can no longer 
be insured then, but the continuity of the fields at $x_2=0$ can. 
The conclusion is that the boundary conditions \eqref{BC} and 
\eqref{layerBelow} lead to the following \textit{cut-off equation},
\be
\label{cutoff}
 \begin{vmatrix}
 \left(c_{44} + \dfrac{e_{24}^2}{\epsilon_{22}}\right)q^2
 + \dfrac{c_{44} \epsilon_{11} + e_{24} e_{15}}{\epsilon_{22}}
   & \hat{c} \left(q^2 + \dfrac{\epsilon_{11}}{\epsilon_{22}}\right) 
  \\[15pt] 
  e_{24} \dfrac{\epsilon_{11}}{\epsilon_{22}} - e_{15}
   & \hat{\epsilon} \left(\dfrac{e_{24}}{\epsilon_{22}}q^2 
                   + \dfrac{e_{15}}{\epsilon_{22}}\right)
 \end{vmatrix}
 = 0.
\en
This equation has an infinity of roots in $kh$ 
(corresponding to the intersections of the graph of $\tan$ with the 
graph of $\coth$). 
Each root  $(k h)_L$ say, is a 
\textit{cut-off parameter} for each dispersion mode, at which the 
Love wave ceases to exist.

Note that the dispersion equation \eqref{theta=0-90}  
is consistent with the secular equation of a 
\textit{Bleustein-Gulyaev wave}
traveling in the substrate alone: as $h$ tends to zero,  
$\hat{c} \rightarrow 0$,  $\hat{\epsilon} \rightarrow \infty$ so that 
it reduces to: 
$\rho v^2 - c_{55} = c_{44}\sqrt{P}$, 
which, once squared, 
coincides with the quadratic \eqref{quadratic}, obtained in the next
Section.
It is also consistent with the dispersion equation for a 
\textit{purely elastic Love wave}. 
Indeed, by taking $e_{ij} \rightarrow 0$ and 
$\epsilon_{ij} \rightarrow 0$ in \eqref{SP} and \eqref{theta=0-90}, 
the equation of Lardat et al. \cite{LaMT71} is recovered:
\be
\tan  \sqrt{\dfrac{\hat{\rho}v^2 - \hat{c}_{44}}{\hat{c}_{44}}}k h 
 = \dfrac{c_{44} \sqrt{\dfrac{c_{55} - \rho v^2}{c_{44}}}}
     {\hat{c}_{44} \sqrt{\dfrac{\hat{\rho}v^2 - \hat{c}_{44}}
                                                  {\hat{c}_{44}}}}.
\en
Finally, it is consistent with the dispersion equation of Love 
surface waves in an isotropic dielectric layer over a 6mm 
piezoelectric substrate \cite{KeLS82}, by the corresponding 
specialization.

\subsection{Rotated cut}

Combining the boundary conditions \eqref{BC} with the results for 
$\hat{t}_{23}(0)$ and $\hat{d}_2(0)$ of \eqref{layerBelow} gives the 
following form for $\vec{\xi}(0)$:
\be \label{cHat}
\vec{\xi}(0) = U_3(0)
      [1,  \alpha, -\ii \hat{c}, -\ii \hat{\epsilon} \alpha]^t,
\en
where $\alpha := \varphi(0)/U_3(0)$ is complex: 
$\alpha = \alpha_1 + \ii \alpha_2$, say.
Then the fundamental equations  \eqref{fundamental} read
\be 
 \begin{bmatrix}
 1 \\ \overline{\alpha} \\ \ii \hat{c} 
     \\ \ii \hat{\epsilon}\overline{\alpha}
 \end{bmatrix}
 \begin{bmatrix}
 M^{(n)}_{11} & M^{(n)}_{12} & M^{(n)}_{13} & M^{(n)}_{14} \\ 
 M^{(n)}_{12} & M^{(n)}_{22} & M^{(n)}_{23} & M^{(n)}_{24} \\ 
 M^{(n)}_{13} & M^{(n)}_{23} & M^{(n)}_{33} & M^{(n)}_{34} \\ 
 M^{(n)}_{14} & M^{(n)}_{24} & M^{(n)}_{34} & M^{(n)}_{44} \\ 
 \end{bmatrix}
 \begin{bmatrix}
 1 \\ \alpha \\ -\ii  \hat{c}  \\ - \ii \hat{\epsilon}\alpha
 \end{bmatrix} = 0,
\en
or
\begin{multline}
 [M^{(n)}_{12} + \hat{c}\hat{\epsilon} M^{(n)}_{34}](2\alpha_1)
  + [\hat{\epsilon}M^{(n)}_{14} - \hat{c}M^{(n)}_{23}](2\alpha_2)
 \\
  + [M^{(n)}_{22} + \hat{\epsilon}^2 M^{(n)}_{44}]
       (\alpha_1^2 + \alpha_2^2) = 
         -[M^{(n)}_{11} + \hat{c}^2 M^{(n)}_{33}].
\end{multline}
Writing them for $n=-1, 1, 2$, and re-arranging the three resulting 
equations, leads to the following non-homogeneous system of linear
equations,
\be \left[ \vec{k_1} | \vec{k_2} | \vec{k_3} \right] \vec{p} 
    = - \vec{k_4},
\en
where $\vec{p} := [2\alpha_1, 2\alpha_2, \alpha_1^2 + \alpha_2^2]^t$
and $\vec{k_1}$, $\vec{k_2}$, $\vec{k_3}$, $\vec{k_4}$ are the
vectors with components:
\begin{multline} 
M^{(n)}_{12} + \hat{c}\hat{\epsilon} M^{(n)}_{34}, \quad 
\hat{\epsilon}M^{(n)}_{14} - \hat{c}M^{(n)}_{23}, \\ 
M^{(n)}_{22} + \hat{\epsilon}^2 M^{(n)}_{44}, \quad 
M^{(n)}_{11} + \hat{c}^2 M^{(n)}_{33}, 
\end{multline} 
($n=-1, 1, 2$) respectively.
Cramer's rule gives the unique solution to the system as
\be \label{solnNonHom}
2\alpha_1 = -\Delta_1/\Delta, \quad 
2\alpha_2 = -\Delta_2/\Delta, \quad 
\alpha_1^2 + \alpha_2^2 = -\Delta_3/\Delta, 
\en
where $\Delta := \text{det}[\vec{k_1}|\vec{k_2}|\vec{k_3}]$, 
$\Delta_1 := \text{det}[\vec{k_4}|\vec{k_2}|\vec{k_3}]$, 
$\Delta_2 := \text{det}[\vec{k_1}|\vec{k_4}|\vec{k_3}]$, and
$\Delta_3 := \text{det}[\vec{k_1}|\vec{k_2}|\vec{k_4}]$. 
The \textit{dispersion equation} follows then from the compatibility 
of the equalities \eqref{solnNonHom}:
\be \label{dispersionGeneral}
\Delta_1^2 + \Delta_2^2 + 4 \Delta_3 \Delta =0.
\en 

When (and if) this dispersion relation yields a real positive 
wave speed $v$ for a 
given wave number $k$, it remains to be checked whether that speed 
corresponds to a valid solution. 
Proceed as follows. 
First recall that the exact boundary condition is of the form 
\eqref{dispersion}, where now the $\vec{a^i}$, $\vec{b^i}$ ($i = 1,2$)
are defined by 
\begin{align}
& \vec{a^i} := 
 \left[ q_i^2 
         + 2\dfrac{\epsilon_{12}}{\epsilon_{22}}q_i
          + \dfrac{\epsilon_{11}}{\epsilon_{22}}, 
         \dfrac{e_{24}}{\epsilon_{22}}q_i^2 
         + 2\dfrac{e_{14}}{\epsilon_{22}}q_i
          + \dfrac{e_{15}}{\epsilon_{22}} \right]^t, 
  \notag \\ 
&  \vec{b^i}:=  (q_i T + R) \vec{a^i}.
\end{align}
The computation of the corresponding surface impedance tensor 
$\ii BA^{-1}$ is long but perfectly possible analytically; 
its components depend on $q_1$, $q_2$ through the sum $q_1+q_2$ and 
the product $q_1q_2$. 
Now, having found a speed from \eqref{dispersionGeneral}, compute 
numerically the roots of the quartic \eqref{quartic}.
Select $q_1$ and $q_2$, the roots with negative 
imaginary parts (if there are no such roots, then $v$ is not valid.) 
Then compute $q_1+q_2$, $q_1q_2$, and $\ii BA^{-1}$, and check 
whether the exact boundary condition \eqref{dispersion} is satisfied.

Finally, it is also possible to determine exactly the \textit{limiting 
speed} $v_L$ at which the imaginary part of an attenuation factor 
first vanishes. 
Fu \cite{Fu05} shows that $v_L$ is the smallest root of $D=0$, where 
$D$ is the discriminant of the cubic resolvent associated with the 
quartic \eqref{quartic}. 
Thus rewrite the quartic  \eqref{quartic} in its canonical form,
\be
p^4 + r p^2 + s p + t =0,
\en
say, using the substitution 
$q = p - (1/2)(\epsilon_{12}/\epsilon_{22})(c_{16}^D/c_{44}^D)$. 
Then, 
\be
D =  -4(12t + r^2)^3 + (-72tr + 2r^3 +27 s^2)^2.
\en
Here the equation  $D=0$ turns out to be a sextic in the squared wave 
speed, according to Maple.
 
\section{Bleustein-Gulyaev wave as $h \rightarrow 0$}

Shuvalov and Every \cite{ShEv02} show that a great variety of 
asymptotic behaviors arises for an interface wave on a coated 
half-space, depending on whether the layer or the substrate is ``fast''
or ``slow'', or ``dense'' or ``light'', depending on which mode is 
under consideration, and depending on several other factors.  

When the thickness of the layer vanishes here, the asymptotic behavior 
of the Love wave in the layer/substrate structure is that of a 
shear-horizontal surface wave propagation over the piezoelectric 
substrate alone (the Bleustein-Gulyaev wave \cite{Bleu68, Guly69}).
For such a wave, with metalized boundary conditions, the vector 
$\vec{\xi}(0)$ takes the form
\begin{equation} 
\vec{\xi}(0) =  [U_3(0), 0, 0, d_3(0)]^t =  U_3(0)[1, 0, 0, \alpha]^t,
\end{equation}
where $\alpha := d_3(0)/U_3(0)$ is complex.
The fundamental equations \eqref{fundamental}, written for 
$n = -1, 1, 2$ (say), can be arranged as
\begin{equation} \label{systemMetal}
\begin{bmatrix}
 M^{(-1)}_{11} &  M^{(-1)}_{14} & M^{(-1)}_{44} \\ 
 M^{(1)}_{11}  &  M^{(1)}_{14 } & M^{(1)}_{44} \\ 
 M^{(2)}_{11}  &  M^{(2)}_{14}  & M^{(2)}_{44} 
\end{bmatrix}
  \begin{bmatrix} 1 \\ 
  \alpha + \overline{\alpha} \\ 
  \alpha \overline{\alpha}
  \end{bmatrix}
= \begin{bmatrix}
      0 \\ 0 \\ 0
   \end{bmatrix},
\end{equation}
a homogeneous linear system of three equations.  
Its solution is non-trivial when the determinant of the 
$3 \times 3$ matrix on the left hand-side is zero.
The resulting secular equation is a \textit{cubic} in $\rho v^2$.
At $\theta =0$ and $\theta = 90^\circ$, the secular equation factorizes
into the product of a term linear in $\rho v^2$ and a term 
quadratic in $\rho v^2$.
In particular, at $\theta = 0$ the quadratic is:
\be \label{quadratic}
(\rho v^2 - \tilde{c}_{55})^2  
  + \tilde{c}_{44}^2
          \dfrac{(\rho v^2 - \tilde{c}_{55})\tilde{\epsilon}_{11}
                    - \tilde{e}_{15}^2}
            {\tilde{c}_{44}\tilde{\epsilon}_{22} + \tilde{e}_{24}^2}
 = 0.
\en
 
\begin{figure}
 \centering 
  \epsfig{figure=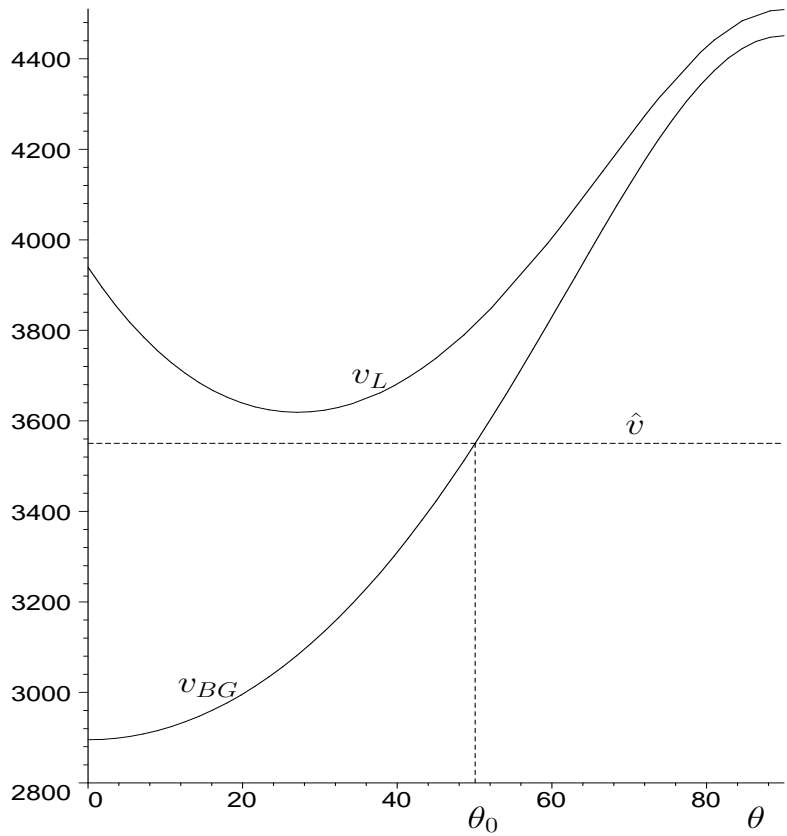, height=.8\textwidth, 
   width=.75\textwidth}
\caption{{\small \sffamily{Solid curves: 
Variations of the Bleustein-Gulyaev wave and limiting wave speeds 
with cut angle in a homogeneous KbNO$_3$ substrate.
Dashed line: Speed of a (bulk) shear wave in germanium.}}}
 \label{Figure2}
\end{figure}

Now for KbNO$_3$, the material parameters of interest are 
\cite{ZSBV93}:
$\tilde{c}_{44} = 7.43$, $\tilde{c}_{55} = 2.5$ (10$^{10}$ N/m$^2$), 
$\tilde{e}_{24} = 11.7$, $\tilde{e}_{15} = 5.16$ (C/m$^2$),
$\tilde{\epsilon}_{11} = 34 \epsilon_0$, 
$\tilde{\epsilon}_{22} = 780 \epsilon_0$ 
($\epsilon_0 = 8.854 \times 10^{-12}$ F/m), 
and $\rho = 4630$ kg/m$^3$.
Using these values, the corresponding parameters in the rotated 
coordinate system follow from  \eqref{c_e_epsilon}, and in turn, 
$T$, $R$, $Q$ follow from \eqref{TRQJ}, $N$ from \eqref{N}-\eqref{Ni}, 
and $M^{(n)}$ from \eqref{fundamental}. 
Then the cubic secular equation is solved for $v$ for any value
of the cut angle $\theta$.
Out of the three possible roots, only one may correspond 
to the Bleustein-Gulyaev wave (Ref.~\cite{CoDe04} explains how 
the adequate root is selected.)
It turns out that the wave exists for all angles, with a speed 
$v_{BG}$ (say) increasing from 
2895.35 m/s at $\theta = 0$ to 4450.85 m/s at $\theta = 90^\circ$.
Fig.\ref{Figure2} shows the dependence in $\theta$, and is in 
agreement with the plot obtained by Nakamura and Oshiki \cite{NaOs97} 
and by Mozhaev and Weihnacht \cite{MoWe00}.

Fig.\ref{Figure2} also displays the speed of the bulk shear wave in a 
layer made of germanium, for which \cite{MaWa05}:
$\hat{c}_{44} = 67.1 \times 10^{10}$ N/m$^2$, 
$\hat{\epsilon}_{11} = 16.6 \epsilon_0$,
and $\hat{\rho} = 5330$ kg/m$^3$; here $\hat{v} = 3550.31$ m/s.
The angle at which $v_{BG} = \hat{v}$ is $\theta_0 = 50.0817^\circ$.
Finally, the variation of the limiting speed $v_L$ in KbNO$_3$ with 
the angle of cut is shown as well, and that plot is also 
in agreement with Mozhaev and Weihnacht \cite{MoWe00}. 

\section{Dispersion curves}

\subsection{Special case $\theta = 0^\circ$}

At $\theta =0^\circ$ and $kh =0$, the interface wave travels in the 
substrate alone, with the Bleustein-Gulyaev wave speed of 2895.35 m/s.
The limiting speed (found here as root of $2\sqrt{P} = S$) is 
$v_L = 3939.33$ m/s. 
The speed of the fundamental mode starts at the Bleustein-Gulyaev wave 
speed at $k h =0$, increases to a maximum speed of about 3857.18 m/s 
at $k h =0.517$, and then decreases toward the shear bulk speed of the 
layer, $\hat{v} =3550.31$ m/s.
In the narrow range where $kh$ is smaller than 
7.08436 $\times 10^{-2}$, the speed $v$ 
of the fundamental mode wave is smaller than $\hat{v}$ and thus 
$\hat{c}$ is given by \eqref{epsilonHat}$_2$; otherwise it is given 
by \eqref{epsilonHat}$_1$. 
The fundamental mode exists for all values of $kh$.

The speeds of the subsequent modes start from the limiting speed $v_L$ 
at $(kh)_L$ and tend to $\hat{v}$ in a monotone decreasing manner. 
The cut-off parameter $(kh)_L$ for the first mode, second mode , 
and third mode is: 6.24, 12.75, and 19.27, respectively. 
Fig.\ref{Figure3} shows the dispersion curves for the fundamental mode 
and for the first mode.
Qualitatively, the plots echo those of Kielczynski et al. \cite{KiPS89}
who considered a structure made of a 6mm substrate covered with a 
``depolarized'' layer.

\subsection{Special case $\theta =90^\circ$}

At $\theta =90^\circ$ and $kh =0$ the limiting speed is root of 
$P = 0$;  here $v_L = 4508.73$ m/s. 
The speed of the fundamental mode starts at the Bleustein-Gulyaev wave 
speed of 4450.85 m/s at $k h =0$, increases to a maximum speed of 
about 4480.69 m/s at $k h =0.0581$, and then decreases toward the 
shear bulk speed of the layer, $\hat{v} =3550.31$ m/s.
Here too, the fundamental mode exists for all values of $kh$.

\begin{figure}
 \centering 
  \epsfig{figure=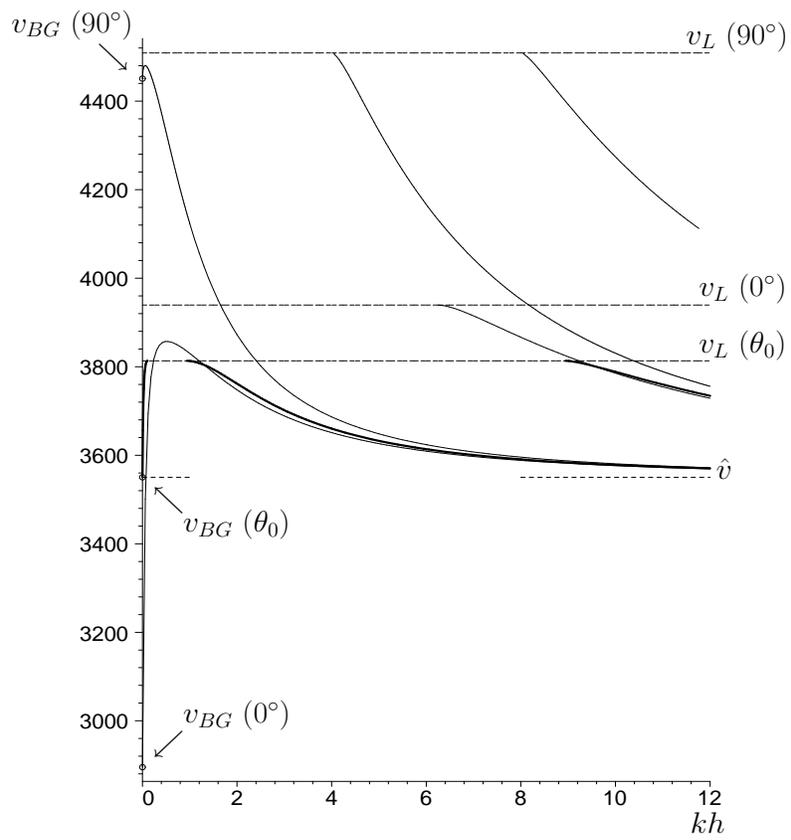, 
       height=.8\textwidth, width=.75\textwidth}
\caption{{\small \sffamily{
Dispersion curves at $\theta=0^\circ, \theta_0, 90^\circ$.}}}
 \label{Figure3}
\end{figure}

The speeds of the subsequent modes start from the limiting speed $v_L$ 
at $(kh)_L$ and decrease toward $\hat{v}$. 
The cut-off parameter $(kh)_L$ for the first mode, second mode , 
and third mode is: 4.05, 8.06, and 12.06, respectively. 
Fig.\ref{Figure3} shows the dispersion curves for the fundamental mode 
and for the first and second modes.

\subsection{Special case $\theta = \theta_0$}
As an example of a rotated cut, consider the case where the speed of 
the Bleustein-Gulyaev wave in the substrate is equal to the shear wave 
speed in the layer; this occurs at $\theta = \theta_0 =  50.08^\circ$,
see Fig.\ref{Figure2}.
For this cut the limiting speed is $v_L = 3813.36$ m/s. 
Starting from $v_{BG} = \hat{v}$ at $kh = 0$, the speed of the 
fundamental mode increases rapidly with $kh$. 
At $kh = 0.1131$, $v = v_L$ and the wave ceases to exists. 
This state of affairs continues until $kh$ reaches 0.9410, after which 
the wave exists and its speed decreases toward $\hat{v}$. 
Hence a \textit{forbidden band of frequencies} emerges for the 
fundamental mode, in clear contrast with the situation for 
non-rotated cuts. 
Note that the dispersion equation \eqref{dispersionGeneral} actually 
gives roots below $v_L$ in that range, which must nevertheless be 
discarded as they do not satisfy the exact boundary condition 
\eqref{dispersion}.

\begin{figure}
 \centering 
  \epsfig{figure=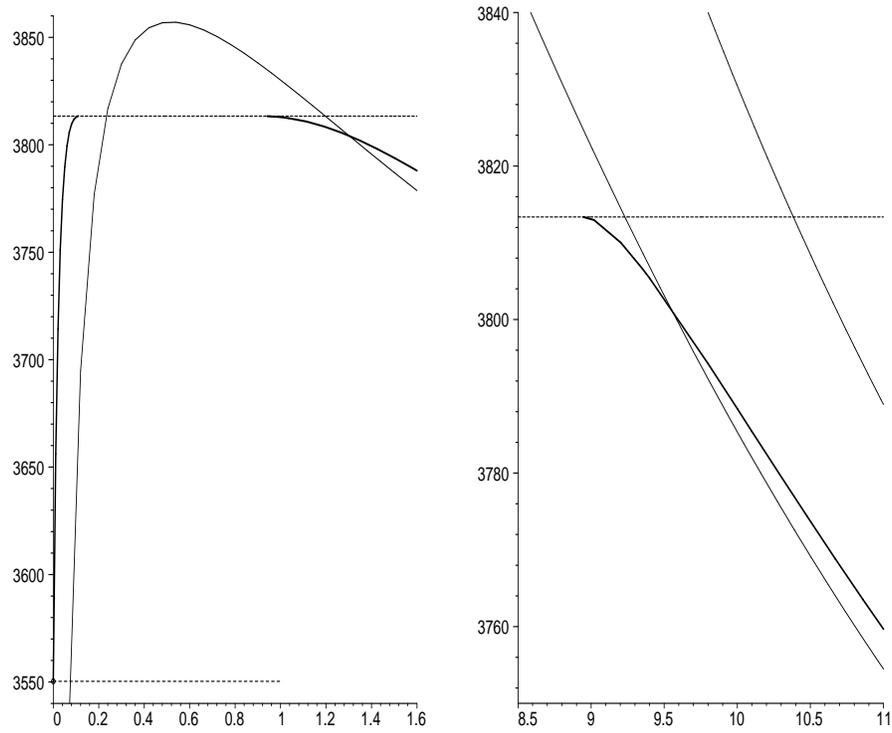, 
       height=.7\textwidth, width=.85\textwidth}
\caption{\small{ 
\sffamily{Zooms for the dispersion of (a) the fundamental mode 
and (b) the first mode at $\theta=\theta_0$ (thin curves) and at 
$\theta = 0^\circ , 90^\circ$ (thick curves) }}.}
 \label{Figure4}
\end{figure}

Here the first mode starts at the cut-off parameter $(kh)_L$ 
of: 8.946 with $v_L$ and then decreases toward $\hat{v}$.
Fig.\ref{Figure4} provides a zoom into the dispersion curves of the 
fundamental mode around the forbidden band and of the first mode up to
 $kh = 12$.

In this example, the layered structure supports a shear-horizontal wave
 which in the long \textit{and} short wavelength ranges travels 
with the speed of the layer's bulk shear wave; 
in the intermediate range, the wave either does not exist, or travels 
at a greater speed.  





\begin{thebibliography}{} 

\bibitem{LaMT71}
  C. Lardat, C. Maerfeld, and P. Tournois,
``Theory and performance of acoustical dispersive surface wave
  delay lines,''
  \textit{Proc. IEEE}, 
  vol.59, pp. 355--368, 1971.

\bibitem{KeLS82}
  G. G. Kessenikh, V. N. Lyubimov, and L. A. Shuvalov,
``Love surface waves in piezoelectrics,''
  \textit{Sov. Phys. Crystallogr.},  
  vol. 27, pp. 267--270, 1982.

\bibitem{HaXi93}
  F. Hanhua and L. Xingjiao,
``Shear-horizontal surface waves in a layered structure of 
  piezoelectric ceramics,''
  \textit{IEEE Trans. Ultrason. Ferro. Freq. Control},   
  vol. 40, pp. 167--170, 1993.

\bibitem{DaWe01}
  A. N. Darinskii and M. Weihnacht,
``Supersonic Love waves in strong piezoelectrics of symmetry $mm2$,''
  \textit{J. Appl. Phys.}, 
  vol. 90, pp. 383--388, 2001.

\bibitem{JaVe97}
  B. Jakoby and M. J. Vellekoop,
``Properties of Love waves: applications in sensors,''
  \textit{Smart Mater. Struct.},  
  vol. 6, pp. 668--679, 1997.

\bibitem{Ogil97}
  J. A. Ogilvy,
``The mass-loading sensitivity of acoustic Love waves biosensors in 
  air,''
  \textit{J. Phys. D: Appl. Phys.}, 
  vol. 30, pp. 2497--2501, 1997.

\bibitem{FaAd72} 
  G. W. Farnell and E. L. Adler,
``Elastic wave propagation in thin layers,'' in
  \textit{Physical Acoustics}, vol. 9,
  W. P. Mason and R. N. Thurston, Eds. 
  New York: Academic Press, 1972, pp. 35--127.

\bibitem{Guly98}
  Yu. V. Gulyaev, 
``Review of shear surface acoustic waves in solids,'' 
  \textit{IEEE Trans. Ultrason. Ferro. Freq. Control},   
  vol. 45, pp. 935--938, 1998.

\bibitem{Auld73}
  B. A. Auld, 
  \textit{Acoustic fields and waves in solids}. 
  Malabar, FL: Krieger, 1990, p. 275. 

\bibitem{MoWe02}
  V. G. Mozhaev and M. Weihnacht,
``Sectors of nonexistence of surface acoustic waves 
  in potassium niobate,''
  in \textit{Proc. IEEE Ultrasonics Symp.}    
  2002, vol. 1, pp. 391--395.

\bibitem{Ting96}
  T. C. T. Ting,
  \textit{Anisotropic elasticity: theory and applications}.
  New York: Oxford University Press, 1996.

 \bibitem{Fu05}
  Y. B. Fu, 
``An explicit expression for the surface-impedance matrix of a 
  generally anisotropic incompressible elastic material in a state of 
  plane strain,'' 
  \textit{Int. J. Non-linear Mech.}, 
  vol. 40, pp. 229--239, 2005.

 \bibitem{DeFu05}
  M. Destrade and Y. B. Fu, 
``The speed of interface waves polarized in a symmetry plane,''
  \textit{Int. J. Eng. Sc.}, 
  vol. 44, pp. 26--36. 2006.

\bibitem{Dest03d}
  M. Destrade,   
``Elastic interface acoustic waves in twinned crystals,''
  \textit{Int. J. Solids Struct.},
  vol. 40, pp. 7375--7383, 2003.

\bibitem{Dest04a}
  M. Destrade, 
``Explicit secular equation for Scholte waves over a monoclinic 
  crystal,''
  \textit{J. Sound Vibr.},
  vol. 273, pp. 409--414, 2004.

\bibitem{Dest05}
  M. Destrade, 
``On interface waves in misoriented pre-stressed 
  incompressible elastic solids,''
  \textit{IMA J. Appl. Math.},
  vol. 70, pp. 3--14, 2005.

\bibitem{Bleu68}
  J. L. Bleustein, 
``A new surface wave in piezoelectric materials,'' 
  \textit{Appl. Phys. Lett.},
  vol. 13, pp. 412--413, 1968.
 
\bibitem{Guly69}
  Yu. V. Gulyaev, 
``Electroacoustic surface waves in piezoelectric materials'', 
  \textit{JETP Lett.},
  vol. 9, pp. 37--38, 1969.
 
\bibitem{ZSBV93}
  M. Zgonik, R. Schlesser, I. Biaggio, E. Voit, J. Tscherry,
  and P. G\"{u}nter, 
``Material constants of KNbO$_3$ relevant for electro- and 
  acousto-optics,'' 
  \textit{J. Appl. Phys.},
  vol. 74, p. 1287, 1993.

\bibitem{CoDe04}
  B. Collet and M. Destrade,  
``Explicit secular equations for piezoacoustic 
  surface waves: Shear-Horizontal modes,''
  \textit{J. Acoust. Soc. Am.},
  vol. 116, pp. 3432--3442, 2004.

\bibitem{NaOs97}
  K. Nakamura and M. Oshiki, 
``Theoretical analysis of horizontal shear mode piezoelectric surface 
  acoustic waves in potassium niobate,''
  \textit{Appl. Phys. Lett.}, 
  vol. 71, pp. 3203--3205, 1997.

\bibitem{MoWe00}
  V. G. Mozhaev and M. Weihnacht, 
``Incredible negative values of effective electromechanical coupling 
  coefficient for surface acoustic waves in piezoelectrics,''
  \textit{Ultrasonics}, 
  vol. 37, pp. 687--691, 2000.

\bibitem{MaWa05}
  W. Martienssen and H. Warlimont, Eds.
  \textit{Springer Handbook of Condensed Matter and Materials Data}.
  Berlin: Springer, 2005.

\bibitem{AbBa90}
  M. Abbudi and D. M. Barnett,
``On the existence of interfacial (Stoneley) waves in bonded
  piezoelectric half-spaces,''
  \textit{Proc. Roy. Soc. London A},
  vol. 429, pp. 587--611, 1990.

\bibitem{ShEv02}
  A. L. Shuvalov and A. G. Every,
``Some properties of surface acoustic waves in anisotropic-coated 
  solids, studied by the impedance method,''
  \textit{Wave Motion},
  vol. 36, pp. 257--273, 2002.

\bibitem{KiPS89}
  P. J. Kielczynski, W. Pajewski, and M. Szalewski,   
``Shear-horizontal surface waves on piezoelectric ceramics with 
  depolarized surface layer,''
  \textit{IEEE Trans. Ultras. Ferroelec. Freq. Control},
  vol. 36,  pp. 287- 293, 1989.
  
  
\end{thebibliography}
\end{document}